# Numerical simulations of gas mixing effect in Electron Cyclotron Resonance Ion Sources

V. Mironov, S. Bogomolov, A. Bondarchenko, A. Efremov, V. Loginov

*Joint Institute for Nuclear Research, Flerov Laboratory of Nuclear Reactions, Dubna, Moscow Reg. 141980, Russia*

The particle-in-cell MCC code NAM-ECRIS is used to simulate the ECRIS plasma sustained in a mixture of Kr with $O_2$, $N_2$, Ar, Ne and He. The model assumes that ions are electrostatically confined in ECR zone by a dip in the plasma potential. Gain in the extracted krypton ion currents is seen for the highest charge states; the gain is maximized when oxygen is used as the mixing gas. A special feature of oxygen is that most of singly charged oxygen ions are produced after dissociative ionization of oxygen molecules with the large kinetic energy release of around 5 eV per ion. Increased loss rate of energetic lowly charged ions of the mixing element requires building up of the retarding potential barrier close to ECR surface to equilibrate electron and ion losses out of the plasma. In the mixed plasmas, the barrier value is large (~1 V) compared to the pure Kr plasma (~0.01 V), with the longer confinement times of krypton ions and with the much higher ion temperatures.

## I. INTRODUCTION

Mixing two gases in ECRIS (Electron Cyclotron Resonance Ion Source) is a common technique to increase currents of the highest charge states of the heavier element. This is the gas mixing effect discovered experimentally by A.G. Drentje [1] in 1983. To see the effect, flow of the lighter gas into a source chamber should be much higher than flow of the working gas. Oxygen is found to be the best mixing gas for such gases as argon, krypton and xenon, with the heavier isotope $^{18}O_2$ giving a mild improvement [2] compared to $^{16}O_2$. Gain in currents depends on the source chamber wall conditions being not pronounced when (oxidized) aluminum is used as the chamber wall material [3,4]. The negative manifestation of the effect is a drastic drop in the extracted ion currents if even very small amounts of the heavier element present in the ECRIS plasma [5].

There are few explanations of what is happening when two gases with different masses are mixed in ECRIS [6]. The most frequently given answer is connected to an evaporative cooling of ions [7]. Ions in the plasma are supposed to be confined within the Electron Cyclotron Resonance (ECR) zone by a negative dip in a globally positive plasma potential [8]. The dip value is such as to equilibrate the ion and electron losses out of the plasma. The ion losses depend on the ion velocities and charge states; the energetic light and lowly charged ion ions are leaving the trap relatively fast, which results in a cooling of those ions that remain be trapped. Then, it is conjectured that in the gas mixed plasmas the ions are colder and thus are better confined by the electrostatic barrier.

At this, no attention is paid to changes in the potential dip value when mixing two gases in the source. Also, it follows from the model that hydrogen and helium should be the best mixing gases, which is not the case. To solve the problem, it is argued that oxygen has higher ionization rates compared to helium, thus giving the higher electron density inside the ECR plasma and the increased rates of ion production for the working gas [9]. Hydrogen as the mixing gas is supposed to be special because of formation of the negative hydrogen ions, which quench the highly charged ions in the charge-change collisions. It is unclear, however, why argon is not effective as the mixing gas for such elements as krypton or xenon, having higher ionization rates compared to oxygen.

Evaporative cooling of ions is believed to be combined with the increased electron life time in the plasma due to decreased electron-ion collision frequency caused by lowering the average ion charge state in the mixed plasma.

D. Meyer et al. [10] argued that the ECRIS plasma is strongly influenced by heavy ions sputtered from the source chamber walls. Fluxes of the sputtered particles depend on energy of ions impinging the walls and, subsequently, on the plasma potential, which is decreasing with addition of light elements into discharge. The gas-mixing effect is considered from its "negative" side as the result of the source performance degradation under an influence of the heavy impurities coming from the walls; authors demonstrated that drop in the extracted currents of nitrogen ions correlates with appearance of copper ions coming from the walls of their copper resonator used as the source chamber. There are doubts, however, whether this works for stainless steel chambers typical for ECRIS and for such working gases as xenon or krypton.

We conclude that exact reasons for the gas-mixing effect remain unknown at the moment. This motivated us to perform the numerical simulations of processes in ECRIS plasma produced in a mix of two different gases. For study of ECRIS, we develop the special code called *NAM-ECRIS* (*N*umerical *A*dvanced *M*odel of *ECRIS*). The results of the calculations show that the ion temperature is actually increased in the gas mixing mode in parallel to increase of the potential dip value. Even with the increased ion temperature, the ratio between the ion confining potential and ion temperature is higher in the gas mixing mode of operation, resulting in the improved ion confinement.

## II. MODEL

The code is based on the model that is described in details elsewhere [11]. The *NAM-ECRIS* is a Particle-in-Cell Monte-Carlo Collisions code that traces a movement of macro-particles representing ions and atoms in ECRIS plasma. Number of macro-particles ($2 \times 10^5$) remains constant during the calculations with the particle statistical weight used as an input. Another input is a temperature of electrons inside the ECR volume ($T_{ew}$), which is varied in the range of few keV. The electron temperature outside zone is always set to 5 eV. The electron density is calculated from charge-neutrality requirement; it is a sum of ion charge densities inside a computational cell.

In calculations, the plasma is characterized with two important values – gas flow in/out of the source chamber and power carried away to the chamber walls by the lost electrons. The values are calculated from the full ion current out of the plasma for the power and from the particle flow into the extraction aperture for the gas flow.

Charged particles undergo elastic and inelastic ion-ion and electron-ion collisions, charge-change collisions with neutral particles and neutralizing collisions with the source chamber walls.

### A. Fields and geometry

Computational particles are moving in the magnetic field of ECRIS. The solenoidal component of the field is calculated with *Poisson/Superfish* code [12]. The multipole component is calculated analytically in the hard-edge approximation. We perform the calculations for the fixed geometry and magnetic field structure of DECRIS-SC2 18 GHz source [13]. The inner diameter of the source chamber is 7.4 cm, the chamber length between the biased disk and extraction electrode is 28 cm, and the chamber is made of stainless steel. Diameter of the extraction aperture is 1 cm. Hexapole magnetic field at the radial wall is 1.1 T, magnetic fields at the axis at the injection and extraction sides of the chamber are 1.97 and 1.35 T respectively. The minimum field is 0.47 T. The magnetic configuration is selected close to the experimentally found optimum for the medium-charged (Q~8+) argon ion production. Calculations are preformed for 18 GHz microwaves resulting in 0.643 T of the electron cyclotron resonance value of the magnetic field.

### B. Potential dip and life times

To see the gas-mixing effect, the model should be modified compared to the version described in [11]. We assume that the ion motion is affected by a dip ($\Delta\varphi$) in the positive plasma potential. The dip or jump in the plasma potential occurs at the ECR surface. The code fixes the moment when an ion crosses the ECR surface; the component of ion velocity along the magnetic field line is calculated. There are two possibilities – either ion moves out of the zone or into the zone. If the ion leaves the ECR volume and its kinetic energy along the line is less than $Q \times \Delta\varphi$ (Q is the ion charge state), the ion is reflected back from the barrier elastically. If the ion is energetic enough to overcome the barrier, its velocity along the magnetic field line is decremented by the corresponding value. When ion moves into the ECR volume from outside, it is accelerated along the magnetic field line with the energy gain $Q \times \Delta\varphi$.

The value of the potential dip $\Delta\varphi$ is selected such as to provide that the calculated ion and electron confinement times in the plasma are equal each with an allowance of ±5%. The ion confinement time is calculated as a ratio between total number of ion charges inside the ECR volume and total ion current ($I_i$) toward the source chamber walls and into the extraction aperture:

$$\tau_i = \frac{\sum_Q (Q \times \int_{ECR} n_{iQ}(x,y,z) dV)}{I_i} \quad (1)$$

The charge-state resolved confinement times of ions are calculated by comparing for each charge state the numbers of ionizing events $I_{ion-Q}$ per second with a flux of ions with charge Q out of the plasma $I_{wall-Q}$:

$$\tau_{iQ}^{-1} = \bar{n}_{eQ} k_Q \frac{I_{wall-Q}}{I_{ion-Q}} \quad (2)$$

Here, an average electron density is calculated for each type of ions during its stay in the hot plasma before ionization; $k_Q$ is the corresponding ionization rate that includes the single and multiple ionization channels. The individual values of the electron density are required to account for the different spatial distributions of different types of ions inside the plasma.

When appropriate, we compare the calculated ion confinement times with the estimation given by Rognlien and Cutler [14] for the collisional ions:

$$\tau_{iQ} = \frac{\sqrt{\pi}RL}{v_i}\exp(\frac{Q\Delta\varphi}{T_i}) \quad (3)$$

Here, L is the length of the system (close to the length of the ECR volume, L=7.3 cm in our conditions), $T_i$ is the ion temperature, R is the mirror ratio ($R = B_{max}/B_{min}$, where $B_{max}$ and $B_{min}$ are the maximal and minimal magnetic fields of the magnetic trap, for this specific case $B_{max}$=0.643 T and the mirror ratio is calculated along the magnetic field lines within the ECR volume, R=1.25) and $v_i = \sqrt{2T_i/M_i}$ is the ion velocity ($M_i$ is the ion mass). The authors of [14] estimate that the times (3) are accurate for the moderately large mirror ratio and barrier height (Q/Δφ≥3$T_i$).

The electron losses are calculated by using the following expression:

$$\nu_e = \tau_e^{-1} = g(R)(\nu_{ei}+\nu_{ee}) + \varepsilon(R,T_{ew})\frac{P_{RF}}{V\langle n_e \rangle T_{ew}} \quad (4)$$
$$+ f(R, E_{sec})\nu_{ion}$$

The first term in the sum accounts for the electron losses into the loss cone due to the electron-electron and electron-ion collisions. The corresponding average 90° scattering frequencies are [15]:

$$\nu_{ee} = 2.9\times 10^{-12}\frac{\langle n_e \rangle}{T_{ew}^{3/2}}\lambda_{ee};$$
$$\nu_{ei} = 4.1\times 10^{-12}\frac{\sum_Q \langle n_{iQ} \rangle \times Q^2}{T_{ew}^{3/2}}\lambda_{ei} \quad (5)$$

Here, electron $\langle n_e \rangle$ and ion densities ($\langle n_{iQ} \rangle$ [m$^{-3}$]) are averaged over the ECR volume; $\lambda_{ee}$ and $\lambda_{ei}$ are the Coulomb logarithms for electron-electron and electron-ion collisions respectively, and $T_{ew}$ is the electron temperature [eV] inside the ECR volume. The g(R) factor in (4) depends on the magnetic trap mirror ratio R. We use the mirror ratio averaged over all magnetic field lines that cross the ECR volume, with taking as the $B_{max}$ value the magnetic field at the point where the line crosses the source wall; R=2.3 for the DECRIS-SC2 18 GHz source.

For g(R) the estimation from R.F. Post [16] is:
$$g(R) = \frac{R+1.5}{R-1} = 2.9 \quad (6a)$$

We note here that the Pastukhov's time [17], which is often used for calculations of electron losses out of magnetic trap of ECR plasma, is derived for R»1 and underestimates the electron loss rate in our case by factor ~3 compared to the Post's time:

$$g(R) = \left[\frac{\sqrt{\pi}}{4}\sqrt{1+\frac{1}{R}}\ln\left[\frac{\sqrt{1+1/R}+1}{\sqrt{1+1/R}-1}\right]\right]^{-1} = 0.8 \quad (6b)$$

The second term in (4) describes the electron losses due to the pitch-angle scattering of electrons by microwaves [18]. There, $P_{RF}$ is the total microwave power absorbed by electrons [eV/sec] and V is the plasma volume [m$^3$], which is supposed to be equal to the ECR volume. For $\varepsilon(R,T_{ew})$ we use the results of the Fokker-Planck calculations of Cluggish et al. [18], fitting them with linear dependence on the electron temperature:

$$\varepsilon(R,T_{ew}) = 0.32\times(\frac{3}{2}T_{ew}/4\cdot 10^4) \quad (7)$$

Cluggish et al. argue that ε should depend on the magnetic trap profile and on the shape of electron distribution function (EDF) in velocity space, independent on the mean electron energy, electron density and microwave power. The fact that the factor ε is increasing with the mean electron energy in their calculations when changing the gas pressure was attributed to be caused by changes in EDF.

The third term f(R, $E_{es}$) in (4) represents the electron losses that occur soon after creation of the secondary electrons in electron-ion collisions [18]. In our model, the new-born electrons are supposed to have an isotropic distribution in velocity space and energies $E_{es}$ equal to ionization potential of the ionized particle [19]. We calculate the probability for the new-born electron to be in the loss cone by saving the starting coordinates of the electrons and their energies for the large number of ionizing events during the calculations; the coordinates and energies are then imported into the special code that traces the electron movement in the source magnetic field. Electrons are supposed to be reflected back from the thin sheath adjacent to the walls if their energy along the magnetic field line is less than 25 eV, which corresponds to the typical value of the plasma potential. Electrons are traced for the sufficiently long time to calculate the number of electrons lost to the walls while bouncing and drifting in the trap.

Electron scattering in collisions with the ions and other electrons is omitted at this stage. The procedure is repeated several times during the calculations to prove that the f(R,E$_{sec}$) value is stable with an accuracy of ±5%. Typical values of the lost electron fraction are in the range 0.05-0.2; the largest values are calculated for krypton because of relatively large energies of the newly created electrons. Without taking into account the electron retardation by the positive plasma potential, the lost electron fraction is 0.3 with no dependence on the electron starting energies.

All factors in (4) are defined with a rather large uncertainty. We use then as the first approximation keeping in mind that separate investigations are needed to calculate the electron losses out of the plasma in more accurate way.

### C. Wall neutralization processes

Gas in the source chamber is heated due to the incomplete energy absorption by a surface after neutralization of energetic ions impinging the walls. We distinguish between the light (lighter than the atoms of wall material) and heavy ions: the heavy ions are supposed to be completely thermalized after their reflection, for the light ions we use the energy accommodation coefficients from [20]. The energy accommodation coefficient is defined as $\alpha = (E_r - E_i)/(E_i - E_w)$. Here, E$_r$ and E$_i$ are energies of the reflected and incident particles respectively, $E_w = \frac{3}{2}kT_w$ is the mean energy of the wall atoms, T$_w$ is the surface temperature. The energy accommodation coefficient depends on angle of incidence of the projectile ($\theta$) and on the ratio between masses of the projectile and wall atoms (u=M$_g$/M$_w$, M$_w$=56), $\alpha = 3.6u\sin\theta/(1+u)^2$. We set the primary energy of ions equal to 25×Q eV (assuming the plasma potential of 25 V), angle of incidence for ions is close to the normal in respect to the surface ($\sin\theta = 1$); for the subsequent collisions of the thermalizing atoms with the walls we use an averaged value for $\overline{\sin\theta} = \sqrt{2}/2$. Ions are supposed to be completely neutralized after their reflection from the surface. For helium, the reflected atoms carry away almost 80% of their primary energy, 20 or 40 eV depending on the ion charge state. Each time as the thermalizing atoms hit the surface, they lose some fraction of their energy and then move slower; time of residence in the source vacuum chamber steadily increases while atoms are cooling. The result is a presence inside the source of the suprathermal atoms with the mean energy of ~0.1 eV.

For the atomic oxygen we take into account a high probability for the atom recombination with forming the molecular oxygen in collisions with the walls ($\alpha_r$~0.5 for the stainless steel surfaces) [21]. We assume that the formed molecular oxygen is fully thermalized after atom recombines in collision with a wall. This probability is relatively small for the atomic nitrogen ($\alpha_r$ ~0.01) [22], as well as for the collisions of atomic oxygen with oxidized surfaces – for the quartz surfaces the coefficient can be as small as $\alpha_r$~10$^{-4}$ [23].

### D. Ionization processes

Ionization rates for the light ions (Z≤30) are taken from the fits of [24]. For the krypton ions, we use the fits from [25] for all charge states except Kr$^0$. For the atomic krypton we use the cross-sections from [26], taking into account large errors in ionization rates of the lowly charged ions in [25]. Scaling from [27] is used for the multiple ionization rates for all gases but argon. For argon, rates for the double ionization are taken from [28]. Ionization and dissociation dynamics of the neutral and singly charged oxygen and nitrogen molecules is treated with taking into account the reactions listed in the Table I. For comparison, ionization rates for the oxygen and nitrogen atoms are also given in the Table I.

It is seen that after dissociation of molecules the singly charged ions and atoms of oxygen and nitrogen are born with the relatively high energies. The oxygen fragments are more energetic compared to the nitrogen ones. The molecular dissociative recombination rates are calculated with the fits from [31, 34]. The rates are non-negligible only for the cold electrons (T$_{ec}$=5 eV) and the recombination is taken into account for the regions outside the ECR volume.

The accepted procedure of the numerical simulations is as follows: we fix the desired level of the coupled microwave power and choose the electron temperature inside the ECR volume. The potential dip value is selected to ensure that the electron and ion confinement times are equal each other; the particle statistical weight is adjusted to reach the selected level of the coupled power.

Table I. Ionization rates (k, $10^{-8}$ cm$^{-3}$/sec) and kinetic energy release (KER) per fragment [eV] for molecular oxygen and nitrogen ($T_{ew}$=12 keV).

|   | Reaction | k | KER | Ref. |
|---|---|---|---|---|
| 1 | $O_2+e \rightarrow O_2^{1+}+2e$ | 5.5 | 0 | 29 |
| 2 | $O_2+e \rightarrow O+O+e$ | 3.3 | 1 | 29 |
| 3 | $O_2+e \rightarrow O+O^{1+}+2e$ | 1.8 | 3.5 | 29 |
| 4 | $O_2^{1+}+e \rightarrow O_2^{2+}+2e \rightarrow O^{1+}+O^{1+}+2e$ | 1.0 | 6.5 | 30 |
| 5 | $O_2^{1+}+e \rightarrow O^{1+}+O^{1+}+2e$ | 1.2 | 6.5 | 30 |
| 6 | $O_2^{1+}+e \rightarrow O^{1+}+O+e$ | 1.8 | 3.5 | 30 |
| 7 | $O_2^{1+}+e \rightarrow O+O$ | 1.0 ($T_{ec}$=5eV) | 1 | 31 |
|   | $O+e \rightarrow O^{1+}+2e$ | 1.4 | 0 | 24 |
| 8 | $N_2+e \rightarrow N_2^{1+}+2e$ | 7.1 | 0 | 32 |
| 9 | $N_2+e \rightarrow N+N+e$ | 3.9 | 0.5 | 32 |
| 10 | $N_2+e \rightarrow N+N^{1+}+2e$ | 0.9 | 3.2 | 32 |
| 11 | $N_2^{1+}+e \rightarrow N_2^{2+}+2e \rightarrow N^{1+}+N^{1+}+2e$ | 1.6 | 5.9 | 33 |
| 12 | $N_2^{1+}+e \rightarrow N^{1+}+N^{1+}+2e$ | 0.6 | 5.9 | 33 |
| 13 | $N_2^{1+}+e \rightarrow N^{1+}+N+e$ | 1.6 | 3.2 | 33 |
| 14 | $N_2^{1+}+e \rightarrow N+N$ | 2.2 ($T_{ec}$=5eV) | 0.5 | 34 |
|   | $N+e \rightarrow N^{1+}+2e$ | 1.3 | 0 | 24 |

## III. RESULTS

A. Injection of one gas

We begin with showing the charge state distributions (CSD) of the extracted ions without mixing the gases. The spectra for krypton plasma are presented in Fig.1 for two electron temperatures $T_{ew}$ (8 and 16 keV) and for the coupled microwave power $P_{RF}$=500 W; spectra for the oxygen plasma are shown in Fig.2 for the same electron temperatures and power. Plasma with the lower electron temperature is obtained by increasing the gas flow if the coupled microwave power is fixed at some level. Increase in the gas flow results in global shift of CSD to the lower charge states with increase in currents of lowly charged ions and with decrease of currents of the highest charge states. This global tendency is often observed in practice.

There is an anomaly in the shape of the krypton CSD at charge state (8+). This is explained by the relatively high ionization rate for Kr$^{8+}$ ions. Other set of ionization rates [35] also shows this anomaly. Measurements confirm the local decrease of the extracted ion currents for this charge state [36].

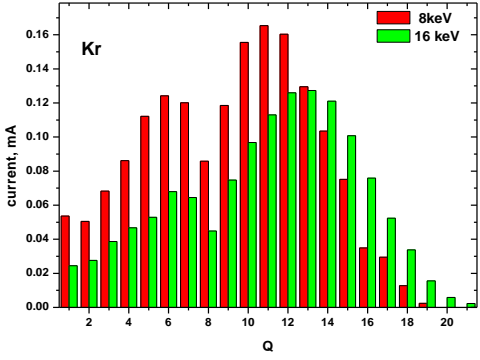

Fig.1. Charge state distribution of extracted krypton ions for the electron temperatures of 8 and 16 keV.

Currents of oxygen ions are much higher compared to the krypton, with the current of $O^{6+}$ reaching 1 mA level. The shape of CSD is close to what is experimentally observed.

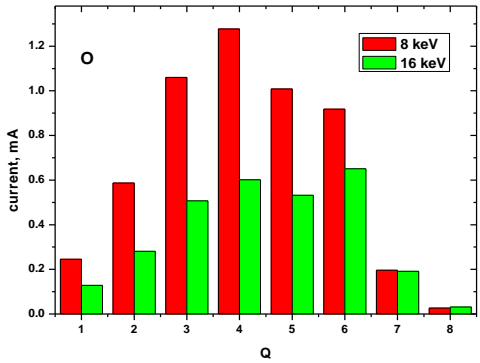

Fig.2. Charge state distribution of extracted oxygen ions for the electron temperatures of 8 and 16 keV.

The calculated parameters of the krypton and oxygen plasma in these conditions (plus the intermediate cases of $T_{ew}$=12 keV and 4 keV) are listed in Table II. Parameters of the plasmas are also shown for $T_{ew}$=4 and 12 keV for injection of helium, neon, argon and nitrogen gases for the same 500 W of the coupled power.

In the Table II, the values are given for the gas flow into the source in particle-mA (for oxygen and nitrogen, flows of the oxygen/nitrogen atoms are given, ×2 of the molecular flow), currents of the extracted ions for the representative charge states, potential dip $\Delta\varphi$, electron confinement time $\tau_e$, ion temperatures for ions inside the ECR volume, and the mean electron density inside the ECR volume.

The largest potential dip values $\Delta\varphi$ are observed for the oxygen plasma, the lowest dips are seen for the krypton plasma. In the descending $\Delta\varphi$ order the elements are sorted as $O_2\rightarrow N_2\rightarrow He\rightarrow Ne\rightarrow Ar\rightarrow Kr$. The ion temperatures follow the same tendency being maximal for the oxygen plasma. Very large difference in the $\Delta\varphi$ values between discharges in the molecular and inert gases is caused by strong heating of oxygen and nitrogen singly charged ions after dissociation of the molecules.

The electron confinement times are largest for He and lowest for Kr, the ordering of elements with the descending confinement times is $He\rightarrow N_2\rightarrow O_2\rightarrow Ne\rightarrow Ar\rightarrow Kr$. The electron density has the same ordering, being maximal for the lightest element in the sequence. We note that the difference in the electron confinement times is not very strong, varying by a factor of around two comparing krypton and helium.

Most of the electron losses from the plasma are caused by the electron-ion scattering process: for krypton, losses due to the electron-electron scattering (Eq.4) equal to around 5% of the total losses, while the RF-induced scattering of electrons contributes to ≈20% of the total losses at $T_{ew}$=12 keV. For helium, the budget of electron losses is as follows: 20% are due to the electron-electron collisions, 20% are caused by the RF-induced loss-cone scattering and the remaining 60% of losses are the result of the electron-ion collisions. In addition, 18% of all new-born electrons are lost soon after their creation for the krypton plasma (factor $f(R,E_{sec})$ in Eq.4). The value for the helium plasma is almost the same, $f(R,E_{sec})$=0.12.

Experimentally, contribution of the RF-induced losses of electrons can be estimated by measuring e.g. electron current to the biased disk after switching off the RF heating of the plasma [37]. Typical drop of the current is around 50%, indicating possible under-estimation of the loss rate in our model. Definitely, more investigation on the subject is needed.

For all investigated gases, increase in the gas flow (decrease in the electron temperature) results in decrease of the electron confinement time and in the lower potential dip values. The electron density is slightly decreasing with increasing the gas flow; changes in the electron confinement times are mainly due to dependence of the electron scattering frequencies on the electron temperature $\sim T_{ew}^{3/2}$, see Eq.4 and 5. The potential dip value drops fast for the krypton and relatively slow for the oxygen plasma. For krypton, the dip value is close to zero at the electron temperature around 8 keV and changes its sign with further increasing the gas flow/decreasing the electron temperature in order to maintain the balance between the electron and ion losses. Neon and argon plasmas show the same tendency, but for them the potential dip approaches zero value at the electron temperature of ~3 keV for the same coupled microwave power of 500 W.

Table II. Main parameters of the plasma with injecting one working gas (krypton, oxygen, helium, neon, argon and nitrogen). The coupled microwave power is 500 W.

| Z | $T_{ew}$, keV | Flow, pmA | $I_i(Q)$, μA | Δφ, V | $\tau_e$, ms | $T_i$, eV | $n_e$, $10^{12}$ cm$^{-3}$ |
|---|---|---|---|---|---|---|---|
| Kr | 4 | 0.77 | 221(12+) | -0.017 | 0.14 | 0.24(12+) | 0.8 |
| Kr | 8 | 0.33 | 160 (12+) | 0.002 | 0.29 | 0.27 (12+) | 0.82 |
| Kr | 12 | 0.25 | 146 (12+) | 0.008 | 0.41 | 0.31 (12+) | 0.81 |
| Kr | 16 | 0.2 | 126 (12+) | 0.012 | 0.55 | 0.32 (12+) | 0.77 |
| O | 4 | 3.0 | 894 (6+) | 0.66 | 0.19 | 3.17 (6+) | 1.06 |
| O | 8 | 1.74 | 918 (6+) | 0.94 | 0.37 | 2.91 (6+) | 1.19 |
| O | 12 | 1.2 | 816 (6+) | 1.1 | 0.54 | 2.85 (6+) | 1.25 |
| O | 16 | 0.9 | 650 (6+) | 1.18 | 0.73 | 2.76 (6+) | 1.25 |
| He | 4 | 8.45 | 7210 (2+) | 0.22 | 0.23 | 0.4 (2+) | 1.52 |
| He | 12 | 3.18 | 3170 (2+) | 0.7 | 0.78 | 0.615 (2+) | 1.75 |
| Ne | 4 | 2.84 | 1320 (6+) | 0.04 | 0.19 | 0.57 (6+) | 1.0 |
| Ne | 12 | 0.88 | 460 (6+) | 0.2 | 0.54 | 0.71 (6+) | 1.15 |
| Ar | 4 | 2.06 | 1450 (8+) | 0.015 | 0.17 | 0.7 (8+) | 0.93 |
| Ar | 12 | 0.6 | 500 (8+) | 0.075 | 0.5 | 0.56 (8+) | 0.97 |
| N | 4 | 3.48 | 2020 (5+) | 0.6 | 0.19 | 2.4 (5+) | 1.17 |
| N | 12 | 1.28 | 991 (5+) | 1.0 | 0.58 | 2.35 (5+) | 1.27 |

The ion temperatures for krypton and other inert gases (except argon) are decreasing with increasing the gas flow into the source, even if the ion heating rate is higher for the lower electron temperatures – the higher heating rate is over-compensated by the decreasing potential dip value and by the decreasing time of ion confinement in the plasma. For oxygen and nitrogen, the ion temperature is increasing with the gas flow: decrease of the dip is not so pronounced for these plasmas and the ion heating rate is mainly determined by dissociation of the molecules.

In contrast to oxygen, currents of the moderately charged nitrogen ions do not saturate with lowering the electron temperature, reaching 2 mA for $N^{5+}$. This difference is due to the relatively faster decrease of the ionization rates for production of highly charged oxygen ions (6+ and higher) at low electron temperatures compared to nitrogen.

B. Injection of two gases

When krypton is mixed with a lighter gas, pronounced gain in currents of the highest charge states of krypton ions can be obtained if flows of the light and main gases are optimized. The typical spectra of extracted ion currents for pure krypton and for the krypton mixed with oxygen are shown in Fig.3. Here, the electron temperature is set to 12 keV, the coupled power is set to 500 W, and the number of oxygen atoms in the source chamber is 85% of the total number of macro-particles. The mix ratio and the electron temperature are selected such as to maximize the extracted $Kr^{18+}$ ion currents. In the mix, currents of Kr ions with the charge states ≥18+ increase, currents of lowly charged ions decrease.

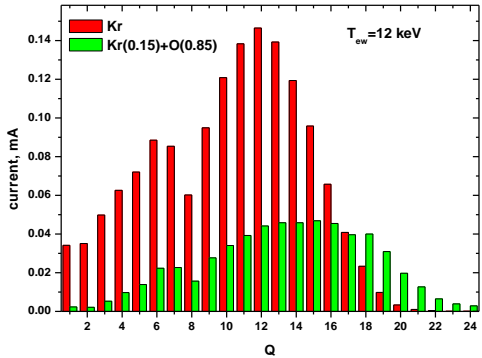

Fig.3. Charge state distribution of the extracted krypton ions in the krypton discharge (red) and in the mix of krypton and oxygen (green).

Dependence of the extracted $Kr^{18+}$ current on the electron temperature in the mix with oxygen (Kr=15% and O=85%) is shown in Fig.4. The coupled power is set to 500 W here, and we mention again that variations in the electron temperature are directly connected to the variations in the gas flow into the source; higher temperature corresponds to the lower gas flow and to the lower total electron/ion fluxes out of the plasma.

As we see in Fig.4, the current of $Kr^{18+}$ ions from the plasma with the above-mentioned mix ratio is maximized at 12 keV, while in the krypton discharge with O=0% this current reaches the maximum at 16 keV.

Maximal currents for the krypton and mixed plasmas differ not so much as when comparing the currents at the same electron temperature of 12 keV. Still, the current of $Kr^{18+}$ in the plasma with the optimized electron temperature and oxygen content is higher by ~15% compared to the maximum in the non-mixed krypton discharge. We note here that the electron losses out of the plasma are calculated without taking into account the losses caused by the plasma micro-instabilities, which may seriously degrade the source performance at the highest electron temperatures: the instability is influenced by the temperature anisotropy along and perpendicular to the magnetic field lines, which increases with the electron temperature. Also, there is a great uncertainty in the rate of electron losses due to RF-induced scattering into the loss cone, the factor ε in Eq.4.

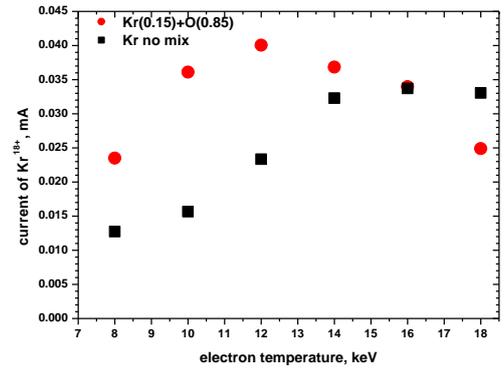

Fig.4. Current of the extracted $Kr^{18+}$ ions as a function of the electron temperature in the krypton discharge and in the mix of krypton and oxygen.

Dependence of the extracted krypton ions on the coupled microwave power is shown in Fig.5. Here, values of the $Kr^{18+}$ current are shown for the krypton plasma (Kr=100%) at the electron temperature of 16 keV. Also, the currents are shown for the mixed plasma (Kr=15%, O=85%) at the electron temperature of 12 keV. Both for the pure krypton and mixed plasmas the current saturates at around 700 W of the coupled power. At the high powers the ion current in the mix substantially exceeds the current from the non-mixed plasma.

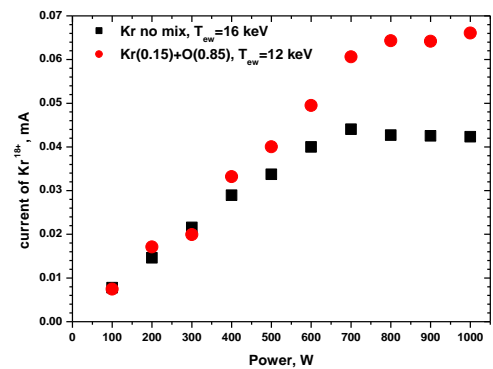

Fig.5. Extracted current of $Kr^{18+}$ ions as a function of the coupled power for the krypton plasma ($T_{ew}$=16 keV) and for the mix with oxygen ($T_{ew}$=12 keV).

The potential dip both for the mixed and non-mixed plasmas varies slowly with the RF power, with the modest increase at low values of the power (<300 W). The electron confinement time is decreasing with $P_{RF}$, reaching the level of 0.34 ms for the mix with oxygen and 0.39 ms for the non-mixed krypton plasma at $P_{RF}$=1000 W. The extracted ion current saturation with the increased power is mainly due to this decrease of the electron confinement time. The relative importance of the RF-induced losses remains constant when comparing the plasmas with the coupled power of 500 W and 1000 W, being at the level of 0.3 of the total electron losses, while the absolute value of the RF loss frequency increases by 30%. The increase of the electron losses is caused by increase both in the electron-ion collision and RF-induced loss frequencies.

In the following, we present the data obtained with the fixed power of 500 W. The selection is a rather arbitrary: the calculated extracted current of $O^{6+}$ ions is at the level of around 1 mA at this power, close to what is measured with the DECRIS-SC2 source when injected microwave power is 600 W. As it is discussed in [11], the calculated value of the coupled power as it is used in our model can substantially differ from the experimentally measured injected power both due to the incomplete microwave absorption in the plasma and deviations of the electron energy distribution function from the Maxwell-Boltzmann one.

*1. Mix with oxygen*

Dependence of the extracted $Kr^{18+}$ ions on the mix ratio is shown in Fig.6. The currents are calculated at the electron temperature of 12 keV. The oxygen content is varied from 0 to 100%. Current of the krypton ions drops by factor of ~2 when small amount (5-10%) of oxygen is added to the discharge, then it grows up and is maximized at 85% of oxygen content. For the oxygen content above the optimal value the current of krypton ions decreases fast.

When changing the oxygen content, gas flows of krypton and oxygen vary almost linearly. The fluxes are shown in Fig.7 for the same plasma parameters as in Fig.6. For the krypton plasma (with no oxygen) the gas flow is 0.2 p-mA, for the oxygen plasma (with no krypton) the flow is 1.2 p-mA. The current of $Kr^{18+}$ is maximized when the oxygen flux is much higher than the krypton flux; the ratio between the fluxes is ~14 for the oxygen content of 85%.

The ion density of krypton ions varies with changing the oxygen content slower than the gas flow. In Fig.8, the mean ion densities of krypton and oxygen ions inside the ECR volume are shown as a function of the oxygen content.

Even for the small krypton content, the mean density of krypton ions is comparable with the density of oxygen ions inside the ECR volume: the ratio between oxygen and krypton densities is 5.3 for the oxygen content of 95%, while the ratio between the gas flows is 63 in these conditions. This is an indication of an increased krypton ion confinement at high oxygen content.

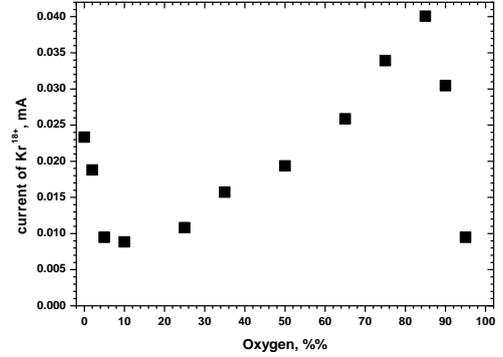

Fig.6. Extracted current of $Kr^{18+}$ ions for different oxygen contents.

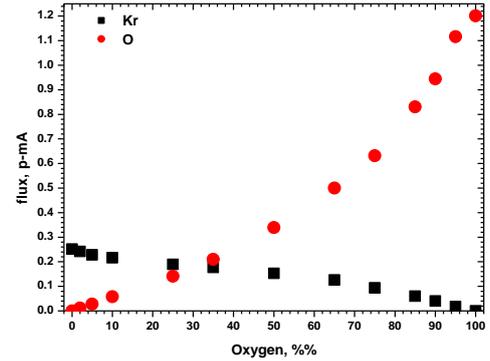

Fig.7. Fluxes of krypton and oxygen atoms into the source for different oxygen content.

Mean charge state of the krypton ions inside the dense parts of the plasma is increasing with increasing the oxygen/krypton mixing ratio. This is illustrated by Fig.9, where the mean charges of krypton and oxygen ions are shown for different oxygen contents. For oxygen ions, the mean charge state does not varies significantly with changing the krypton content in the wide range down to Kr=5% being at the level of ~(2+), much lower compared to the pure oxygen plasma (Kr=0%), for which it is close to (4+). For the krypton ions, their mean charge state is increasing when adding more oxygen into the plasma and reaches (14+) at Kr=5%, almost doubling compared to the pure krypton plasma case.

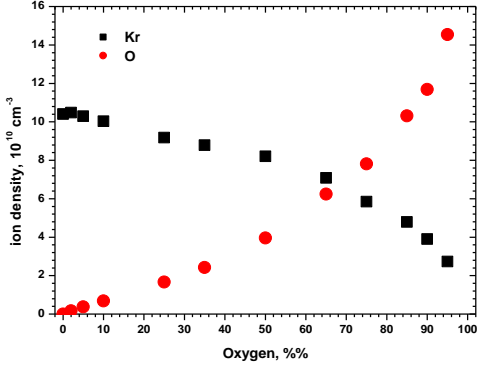

Fig.8. Total densities of krypton and oxygen ions averaged over the ECR volume for different oxygen contents.

The mean electron density inside the ECR volume is not changing significantly for different oxygen mixings, being at the level of $8 \cdot 10^{11}$ cm$^{-3}$. The only change is a fast increase in the density in the O=100% case. There, the electron density is noticeably higher and reaches $1.25 \cdot 10^{12}$ cm$^{-3}$ level (see also Table II).

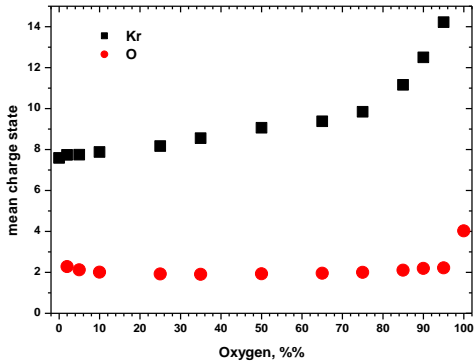

F

Fig.9. Mean charge state of the krypton and oxygen ions inside the ECR volume for different oxygen contents.

The electron confinement time is at the level of 0.4 ms for all plasmas with the non-zero krypton content. Again, there is a jump in the electron confinement time in the O=100% case, for which it is increased up to 0.54 ms.

For all investigated mixes, the electron confinement time is mostly determined by the electron-ion scattering; for the non-zero krypton contents the contribution of electron scattering on the krypton ions into the total scattering frequency is determinative. Even for the limiting krypton content Kr=5%, when the krypton flux into the source is much lower than the flux of oxygen, the frequency of electron-krypton collisions is ~0.9 of the total electron-ion scattering frequency. As it is following from Eq.5, the electron-ion frequency scales as ~$Q^2$, where Q is an ion charge state. Even with having the relatively small densities, the krypton ions scatter the plasma electrons more frequently because of their high charge states.

The increased oxygen content leads both to increase of the potential dip value and to increase of the ion temperatures. Dependencies of $\Delta\varphi$ and temperature of $Kr^{17+}$ ions inside the ECR volume are shown in Fig.10, as well as a ratio between the dip and ion temperature $\Delta\varphi/T_i(Kr^{17+})$. The charge state (17+) is selected because these ions are a source for production of $Kr^{18+}$ ions and we are mainly focused on the extracted currents of $Kr^{18+}$ ions in our analysis.

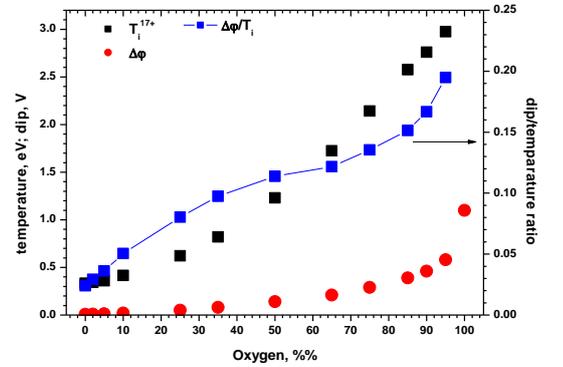

Fig.10. Potential dip $\Delta\varphi$ (red circles, left scale), temperature of $Kr^{17+}$ ions (black squares, left scale) and the ratio between these values (blue squares, right scale) for different oxygen contents.

The dip is growing faster than the ion temperatures resulting in the increasing ratio $\Delta\varphi/T_i(Kr^{17+})$ and in stronger ion confinement.

More details of changes in ion confinement are given in Fig.11, where the confinement time of $Kr^{17+}$ ions is plotted as a function of the oxygen content. The time is calculated by using Eq.2; fast increase in the confinement time is seen. The time is increased by a factor of almost three at the optimized mix of O=85% compared to the krypton plasma. It is also seen that injection of small amount of oxygen results in a decrease of ion confinement time by around 30%.

Dependence of the ion confinement time is fitted in Fig.11 with the "Ronglien-Cutler"-type curve:

$$\tau_{iQ} = \frac{A}{9.79 \cdot 10^3 \sqrt{2T_{iQ}/M_i}} \exp(Q\Delta\varphi/T_{iQ})$$

where A is the fitting coefficient [m], $M_i$ is the ion mass in atomic units and $9.79 \cdot 10^3$ m/sec is the unit conversion factor. The fitting coefficient of the curve in Fig.9 is A=0.68 m. The estimate for Eq.3 gives A=0.16 for R=1.25 and L=0.073 m; the times in Fig.9 correspond to $\approx 4.25\tau_{(Ronglien-Cutler)}$ for the high and very small oxygen contents.

Strong deviations from the fitting curve are seen for the low and intermediate oxygen contents in the range from 5 to 65%.

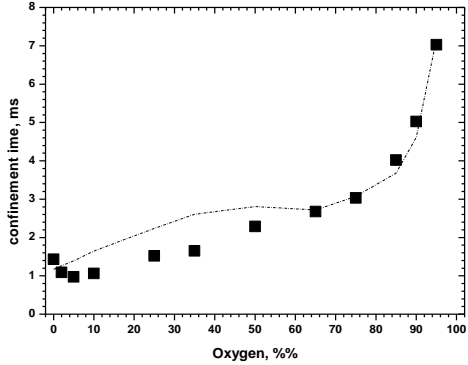

Fig.11. Confinement time of $Kr^{17+}$ ions for different oxygen contents.

The drop in the extracted krypton ions (Fig.6) and decreased ion confinement times at low oxygen content in the interval O=(5-25)% can be understood by analyzing the ion pressure profiles.

In Fig.12, dependencies of ion pressure on z-coordinate along the source axis are shown for different oxygen contents of 0 and 5%. The profiles are calculated as $P(z) = \sum_i n_{iQ}(z) T_{iQ}(z)$, where summation is done for all ions of the specific element, krypton or oxygen. The dashed lines indicate the ECR zone positions. For all plasmas, the ion pressure decreases fast outside the ECR zone both in directions to the injection and extraction sides of the source (injection side is at z=0). For the krypton plasma (O=0%) the profile is a rather flat inside the ECR zone, while for the relatively low oxygen content of 5% the profile of krypton ion pressure is hollow at the source center. At the same time, oxygen ion pressure is peaked at the center.

Ion density of oxygen is small compared to the density of krypton ions at this small oxygen content, but the oxygen pressure is comparable to the krypton pressure because of high energies of the oxygen ions. The oxygen ions push the krypton ions toward the ECR zone boundaries, degrading their confinement. When the oxygen content is high, potential dip starts to be large enough to retard most of the energetic oxygen ions; the oxygen pressure profile inside the ECR volume becomes be flat and the loss of krypton ion confinement diminishes.

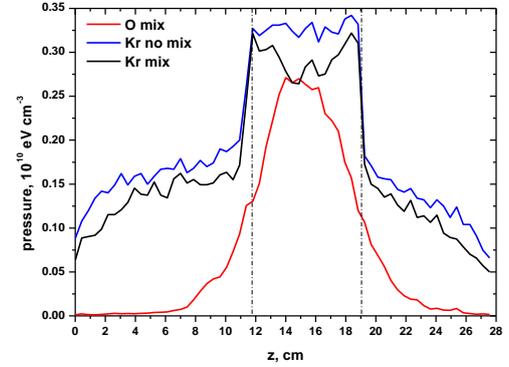

Fig.12. Ion pressure profiles along the source z-axis: pressure of the oxygen ions (red), pressure of the krypton ions at Kr=95%, O=5% (black), pressure of the krypton ions at Kr=100% (blue). $T_{ew}$=12 keV, $P_{RF}$=500 W.

We are not aware about the direct experimental measurements of how small amounts of oxygen influence the source output for the heavy elements. The indirect confirmation of the effect can be that any ECRIS should be conditioned after breaking the vacuum to reach a good source performance. Apart from changing the source chamber wall conditions, the source conditioning can be connected with the process of removing the residual oxygen and nitrogen molecules out of the source.

We see from Fig.10 that the potential dip value decreases substantially when small flux of krypton atoms is injected into the plasma – krypton content Kr=5% leads to the Δφ decrease from 1.1 to 0.58 V. The result is a loss of confinement for oxygen ions and decrease in the extracted oxygen currents for the high charge states. In Fig.13, the charge state distributions for oxygen are shown for oxygen plasma O=100% and for the krypton-oxygen mix Kr=5%, O=95%. The current of $O^{6+}$ decreases by an order of magnitude.

Ion temperatures are not the same for different charge states of ions. The general tendency is an increase of the ion temperature with the ion charge state, especially pronounced in the mixed plasmas. Dependences of the temperatures on the charge state are shown in Fig.14 for the krypton plasma Kr=100% and for the mix Kr=15%, O=85%. For the krypton plasma, the ion temperatures are changing by a factor of two comparing the lowly charged (1+) and highly charged (20+) ions. For the mixed plasma, the span in the ion temperatures is much higher, with the temperature of $Kr^{1+}$ ions of around 0.2 eV and of 3 eV for $Kr^{20+}$. This is an indication of different ion confinement times and different rates of ion energy changes in electron-ion/ion-ion collisions.

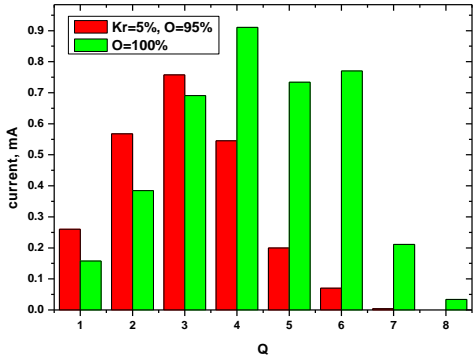

Fig.13. Charge state distribution of extracted oxygen ion currents for oxygen plasma and for the mix with krypton (Kr=5%, O=95%). Electron temperature $T_{ew}$=12 keV, $P_{RF}$=500 W.

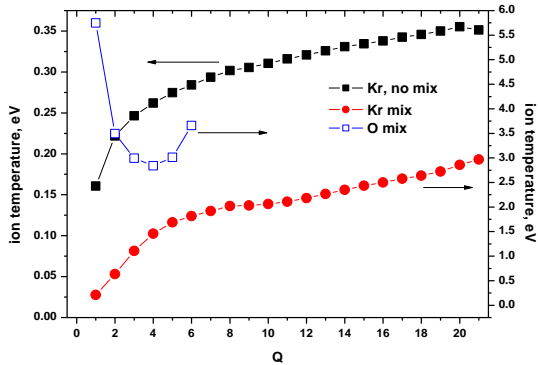

Fig.14. Charge-state dependence of the ion temperature for krypton ions in the pure krypton discharge (left scale, solid black squares) and in the mix of krypton and oxygen, (right scale, red circles). Temperatures of oxygen ions are shown as open blue squares (right scale).

Charge state dependences of the ion confinement times are shown in Fig.15 for the krypton and mixed plasmas.

Confinement time increases with the ion charge state. For the krypton plasma, saturation in the dependence is seen for the high charge states above 10+. In the mix, confinement times of krypton ions are much higher compared to the non-mixed krypton plasma; the gain is around factor of ~4. In the same conditions, confinement times of oxygen ions is much smaller than the times for the krypton ions with the same charge states, reflecting the higher temperatures and the higher mobility of oxygen ions.

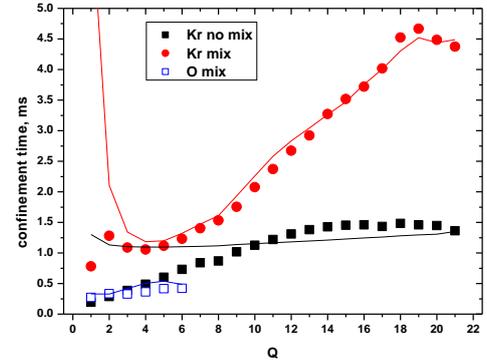

F

Fig.15. Confinement times of krypton ions for the krypton plasma (black squares) and in the mix with oxygen (red circles) as a function of the ion charge state. Confinement times of oxygen ions in the mix are shown as the open blue squares.

The dependencies are fitted with the Rognlien-Cutler type curves as in Fig.11. The fits are shown in Fig.15 as the lines. The fitting coefficient is the same for all curves, A=0.74 m, close to what is calculated for the dependence in Fig.11. It is seen that the fit greatly over-estimates the confinement times for the lowly charged ions both in the mixed (Q~1+-2+) and non-mixed plasmas (Q<10+). Good correspondence between the calculated times and the fit is obtained for the highly charged krypton ions in the mixed plasma.

*2. Mixes with other gases*

Mixed plasma parameters are also obtained for other mixing gases (N, He, Ne and Ar), and for the oxygen isotope $^{18}O$. The results are listed in the Table III, showing the extracted currents of $Kr^{18+}$, flows of the krypton and mixing gases, currents of the representative ions of the mixing element, potential dip value, the electron confinement time, temperature of $Kr^{17+}$ ions in the ECR volume, temperature of the representative ions of the mixing element, mean electron density in the ECR volume, electron density seen by $Kr^{17+}$ ions, and the confinement time of the $Kr^{17+}$ ions.

Table III. Main parameters of the krypton plasmas mixed with different gases. Krypton content is Kr=15%, $T_{ew}$=12 keV, $P_{RF}$=500 W.

| Z | Flow (Kr), pmA | Flow (mix), pmA | $I_i$ (Kr$^{18+}$), μA | $I_i$(Q), μA | Δφ, V | $\tau_e$, ms | $T_i$(17+), eV | $T_i$(Q), eV | $n_e$, $10^{12}$ cm$^{-3}$ | $n_e$(Kr$^{17+}$), $10^{12}$ cm$^{-3}$ | $\tau_i$(Kr$^{17+}$), ms |
|---|---|---|---|---|---|---|---|---|---|---|---|
| $^{16}$O | 0.06 | 0.83 | 40 | 28(6+) | 0.39 | 0.46 | 2.64 | 3.66(6+) | 0.73 | 0.44 | 4.0 |
| $^{18}$O | 0.06 | 0.79 | 38.5 | 32(6+) | 0.39 | 0.47 | 2.68 | 3.52(6+) | 0.72 | 0.45 | 3.9 |
| N | 0.054 | 0.87 | 33 | 63(5+) | 0.3 | 0.48 | 1.93 | 2.48(5+) | 0.69 | 0.40 | 5.1 |
| He | 0.074 | 1.23 | 12.4 | 285(2+) | 0.022 | 0.45 | 0.2 | 0.21(2+) | 0.75 | 0.79 | 2.6 |
| Ne | 0.051 | 0.70 | 19.9 | 198(6+) | 0.05 | 0.50 | 0.36 | 0.38(6+) | 0.83 | 0.57 | 3.2 |
| Ar | 0.047 | 0.48 | 13.4 | 342(8+) | 0.06 | 0.48 | 0.55 | 0.54(8+) | 0.89 | 0.62 | 2.4 |

There is no statistically significant difference between mixed Kr-O plasmas with injection of light $^{16}$O and heavy $^{18}$O isotopes. When using nitrogen as the mix gas, current of Kr$^{18+}$ is smaller than in the oxygen mix. Also, both potential dip and the ion temperatures are smaller. The ion confinement time for Kr$^{17+}$ ions with the admixed nitrogen is higher than in oxygen mix, as well as the ratio between the Δφ/$T_i$(Kr$^{17+}$) values (0.155 for the nitrogen and 0.147 for the oxygen mix). Ions are colder in the Kr-N plasma because of two main reasons: smaller kinetic energy release after ionization of nitrogen molecules and larger contribution of the suprathermal nitrogen atoms into the production of the lowly charged nitrogen ions in the dense parts of the ECR plasma. The ions that are produced from the suprathermal nitrogen atoms have the relatively low energies and cool the ion population. For nitrogen we use small recombination coefficient for production the molecular nitrogen after atom collisions with the walls. Calculations with the same recombination coefficient as for oxygen (0.5) give the potential dip value of 0.37 V comparable with the value for the oxygen mix. Still, the extracted Kr$^{18+}$ current is smaller in these conditions compared to the oxygen mix.

What makes the oxygen-mixed plasma more efficient for production and extraction of the highly charged ions of krypton is the spatial distribution of the ion densities, which is hollow but more concentrated toward the source axis compared to the nitrogen case. Indeed, in the oxygen plasma the mean electron density seen by the krypton highly charged ions is larger by ~10% compared to the nitrogen plasma. The ion distribution at the extraction electrode is more peaked at the source axis - more ions pass through the extraction aperture. Ion densities at the middle of the source along x-axis are plotted in Fig.16 for oxygen and nitrogen mixes, as well as for the non-mixed krypton plasma, for all krypton ions with Q≥17+.

The plasma spatial profile is defined by the ambipolar diffusion of particles across the magnetic field due to the electron-ion collisions and by the spatial diffusion of ions caused by the unlike elastic ion-ion collisions [11]. The plasma shape depends, among other factors, on the spatial gradients of the magnetic field, plasma composition and ion temperatures. Hotter ions in the oxygen-krypton plasma make the profile broader than in the relatively colder nitrogen-krypton mix. For the non-mixed krypton plasma, the profile is the sharpest and the ion densities are smallest compared to the mixed plasmas.

We note here that the densities are connected to the extracted ion currents with a scaling factor equal to the ion confinement time ($I_i \sim n_i/\tau_i$) and the ion confinement times are smaller for the non-mixed plasma. The extracted ion currents differ not so much as the ion densities when comparing the mixed and non-mixed plasmas.

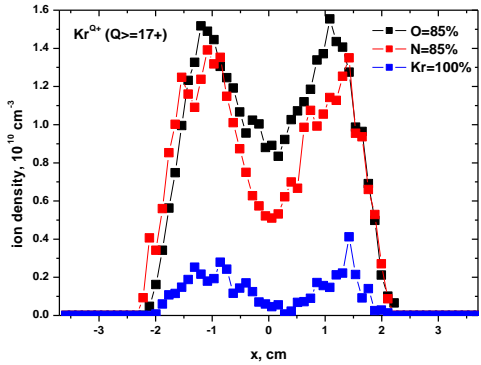

Fig.16. Density of the krypton ions with the charge states greater and equal to (17+) along x-axis in the middle of the source (z=14 cm) for the mix with oxygen (O=85%, black), nitrogen (N=85%, red) and with no mix (Kr=100%, blue).

Neon is the best among the mixing noble gases, still providing much smaller currents of krypton highly charged ions and smaller potential dip values than in the oxygen and nitrogen mixes. Argon and helium are less effective as the mixing gases compared to neon, with the helium mix resulting in the smallest potential dip values and smallest current of $Kr^{18+}$ ions.

## IV. CONCLUSIONS

Combination of the three-dimensional calculations of ion dynamics in the ECRIS plasma and the plasma-averaged calculations of electron confinement times allows reproducing the plasma parameters both in the single-gas and gas-mixed discharges. The gas mixing effect is seen for mixing krypton with some lighter gases; the highest gains in the currents of the highly charged krypton ions are for the mix with oxygen. The reasons for the effect are due to increase of the potential dip that confines the ions inside the dense parts of the ECRIS plasma. Ionization of oxygen and nitrogen molecules results in the energization of the singly charged ions produced after the molecule dissociation. Temperature of krypton ions increases in the mixed plasma because of the extra heating by the energetic lowly charged ions of the mixing gas, improved ion confinement and boost in the heating rate by the electron-ion collisions with the increased mean charge state of ions. Changes in the spatial distribution of ions in the plasma are seen in the mix. Drop of the highly charged ion currents of the lighter element is observed when adding small fluxes of krypton. The drop is caused by accumulation of the krypton ions inside the plasma, which decreases the potential dip and the electron/ion confinement times.